\documentclass[aps,pre,twocolumn,showpacs,superscriptaddress]{revtex4}
\usepackage{epsfig,amsmath,amssymb,bm,epsf,graphics,graphicx,verbatim,subfigure,color}
\usepackage{slashed}
\usepackage{float}
\usepackage{graphicx}
\usepackage{pdfpages}

\def\nzm{N_{\textrm{zm}}} 
\def\nss{N_{\textrm{ss}}} 
\def\ndof{N_{\textrm{d.o.f.}}} 
\def\nldof{n_{\textrm{d.o.f.}}} 
\def\ncon{N_{\textrm{con}}} 
\def\nlcon{n_{\textrm{con}}} 
\def\rmat{\mathbf{C}} 
\def\qmat{\mathbf{Q}} 
\def\uvec{\mathbf{u}} 
\def\evec{\mathbf{e}} 
\def\ehat{\hat{e}} 
\def\fvec{\mathbf{f}} 
\def\dim{d} 
\def\nvec{\mathbf{n}} 
\def\zvec{\mathbf{z}} 
\def\en{U} 
\def\sh{s} 
\def\ind{n} 
\def\chart{\chi_\textrm{T}} 
\def\reg{\mathcal{B}} 
\def\mdeg{n} 
\def\jac{J_{C}} 
\def\area{A} 
\def\nmm{N_{--}} 
\def\nmp{N_{-+}} 
\def\npm{N_{+-}} 
\def\npp{N_{++}} 
\def\springs{\mathbf{K}} 
\def\ra{\mathbf{r_1}} 
\def\rb{\mathbf{r_2}}
\def\rc{\mathbf{r_3}}
\def\rd{\mathbf{r_4}}
\def\ta{\phi_1} 
\def\tb{\phi_2}
\def\tc{\phi_3} 
\def\lc{\mathbf{s}} 
\def\pvec{\mathbf{\phi}} 
\def\hatf{\hat{f}} 
\def\vvec{\mathbf{v}} 
\def\rotm{\theta} 

\begin{document}

\title{Controlling the Deformation of Metamaterials: Corner Modes via Topology}

\author{Adrien Saremi}
\affiliation{School of Physics, Georgia Institute of Technology, Atlanta, GA 30332}

\author{Zeb Rocklin}
\affiliation{School of Physics, Georgia Institute of Technology, Atlanta, GA 30332}

\bigskip

\begin{abstract}
Topological metamaterials have invaded the mechanical world, demonstrating acoustic cloaking and waveguiding at finite frequencies and variable, tunable elastic response at zero frequency. Zero frequency topological states have previously relied on the Maxwell condition, namely that the system has equal numbers of degrees of freedom and constraints. Here, we show that otherwise rigid periodic mechanical structures are described by a map with a nontrivial topological degree (a generalization of the winding number introduced by Kane and Lubensky) that creates, directs and protects modes on their boundaries. We introduce a model system consisting of rigid quadrilaterals connected via free hinges at their corners in a checkerboard pattern. This bulk structure generates a topological linear deformation mode exponentially localized in one corner, as investigated numerically and via experimental prototype. Unlike the Maxwell lattices, these structures select a single desired mode, which controls variable stiffness and mechanical amplification that can be incorporated into devices at any scale.
\end{abstract}

\maketitle

\section{Introduction}

Topological phases of matter have been realized most famously in 
electronic systems~\cite{thouless1982quantized,kane2005quantum,bernevig2006quantum,teo2010topological,hasan2010colloquium,qi2011topological},
 but also in classical ones consisting of active fluids~\cite{van2016spatiotemporal,souslov2017topological,murugan2017topologically,dasbiswas2017topological}, 
air flows~\cite{khanikaev2015topologically,he2016acoustic,chen2016tunable},
 photons~\cite{rechtsman2013photonic, lu2014topological,peano2015topological},
 vibrating mechanical elements~\cite{susstrunk2015observation,vila2017observation,trainiti2018optical} 
and spinning gyroscopes~\cite{nash2015topological,wang2015topological,mitchell2018amorphous}.
 Across this vast range of systems, the topological paradigm: 1) identifies various invariants that assume discrete values determined by the bulk structure that 2) are insensitive to continuous deformations and 3) determine protected edge modes. We consider yet another class of topological systems, zero-frequency mechanical ones, that has been shown to have a topological invariant protected by the Maxwell condition, that the system's degrees of freedom (d.o.f.) and constraints are equal in number, and that determines the placement of topological modes on open boundaries and interfaces~\cite{kane2014topological}, point defects~\cite{paulose2015topological}, and even the bulk~\cite{rocklin2016mechanical}. Such systems have been demonstrated or proposed as 
new ways of controlling origami and kirigami folding~\cite{chen2016topological}, beam buckling~\cite{paulose2015selective} and fracture~\cite{zhang2018fracturing}, as well as composing 
nonreciprocal mechanical diodes~\cite{coulais2017static} and
mechanically programmable materials~\cite{paulose2015topological}. Their acoustic counterparts have wave propagation that is backscattering free~\cite{
susstrunk2015observation,
pal2016helical,vila2017observation} and, when time reversal symmetry is broken, unidirectional
~\cite{khanikaev2015topologically,wang2015coriolis,wang2015topological,nash2015topological,mitchell2018amorphous}, permitting unprecedented waveguiding and cloaking capabilities.

As well as such examples of topological modes one dimension lower than the bulk, exciting new predictions have been made concerning higher-order topological modes on lower-dimensional surface elements, such as those split between the four corners of a square 2D system
~\cite{benalcazar2017quantized,benalcazar2017electric,PhysRevLett.119.246402,PhysRevLett.119.246401,schindler2017higher}. Such modes, subsequently observed in a phononic system~\cite{serra2018observation} and a microwave circuit system~\cite{peterson2018quantized} with mirror symmetries, raise the possibility of higher-order mechanical modes. Here, we report just such a family of topological lattices, presenting both a general theory and a detailed and experimentally realized example: the 2D ``deformed checkerboard'' lattice. 
These lattices possess higher-order mechanical criticality, in the sense of having modes localized to lower-dimensional sections of their boundaries.
In contrast to the topological polarization of Maxwell lattices, isolated zero modes are present in otherwise \emph{rigid} materials (and the force-bearing self stresses in otherwise floppy ones), amounting to a fundamentally new and topologically nontrivial capability among flexible mechanical metamaterials~\cite{bertoldi2017flexible}.

The remainder of our paper is arranged as follows. 
In Sec.~\ref{sec:hom}, we introduce higher-order Maxwell rigidity and describe its topological properties.
In Sec.~\ref{sec:check} we illustrate the family of structures with a single, experimentally realized example. Finally, in Sec.~\ref{sec:discussion} we discuss the implications for future work.

\section{Higher-order Maxwell rigidity}
\label{sec:hom}

\subsection{A new counting argument}

Consider a system governed by a \emph{constraint matrix} $\rmat$ which linearly maps some coordinates $\uvec$ (often displacements of sites) to another vector $\evec$ (often extensions of stiff mechanical elements). Modes $\evec$ in the nullspace of $\rmat^{\textrm{T}}$ are called \emph{self stresses}, (generalized) tensions that do not generate force. Because of this relation, the rank-nullity relation of linear algebra implies

\begin{align}
\label{eq:index}
\nzm-\nss = \ndof - \ncon,
\end{align}

\noindent where the four symbols refer respectively to the system's numbers of zero modes, self stresses, degrees of freedom and constraints. This equality is, in the context of constraint matrices of ball and spring systems, owed to Calladine~\cite{calladine1978buckminster}, Maxwell a century earlier noting the phenomenon of redundant constraints, but not identifying self stresses~\cite{maxwell1864calculation}. We consider a generalized notion of constraints leading to generalized pseudo-forces.

For periodic systems in $\dim$ dimensions, this can be further refined by assuming that the displacements within a single cell indexed by 
$\nvec = \left(\ind_1,\ind_2,\ldots \ind_d \right)$ are of the ``z-periodic''  form $\uvec_\nvec = \uvec \prod_i z_i^{\ind_i}$, where $z_i$ is a complex number of any magnitude, as used in surface quantum wavefunctions~\cite{PhysRevLett.117.076804,PhysRevB.96.195133,cobanera2017exact}. Indeed, due to an argument similar to Bloch's theorem, any normal mode of a system with a periodic bulk must have such a form.
For a finite system with open boundary conditions, a mode with, e.g., $|z_1|<1$ exists exponentially localized on the left-hand boundary. Kane and Lubensky exploited this index theorem in \emph{Maxwell} systems (those with $\nldof = \nlcon$), to relate the number of such zero modes, for which $\det[\rmat(\zvec)]=0$, to the winding of the phase of $\det(\rmat(\zvec))$ as $z_1$ winds around the bulk ($z_1 = e^{i \theta_1}$)~\cite{kane2014topological}. As we now present, this one-dimensional winding number is only one example of a whole family of topological invariants.

The Maxwell condition is a mechanical critical condition~\cite{lubensky2015phonons} which identifies systems expected to have boundary modes from missing bonds at boundaries. More generally, rather than fixing some inter-cell evolution via $\zvec$, we can vary $\mdeg$ of the $z_i$ ($\mdeg \le \dim$) as well as $\uvec$, the shape of the mode within a single cell, generating $\mdeg$ additional ``degrees of freedom''.
Thus, upon normalizing our linear mode, the configuration space has dimension $\mdeg +\nldof -1$, and zero modes require satisfying $\nlcon$ constraint conditions. Thus, the critical condition under which we would expect (for compatible and independent constraints) isolated zero modes on a surface element $\mdeg$ dimensions lower than the system's bulk is

\begin{align}
\label{eq:ourindex}
\mdeg = \nlcon + 1 - \nldof.
\end{align}

\begin{figure}
\center
\includegraphics[scale=0.5]{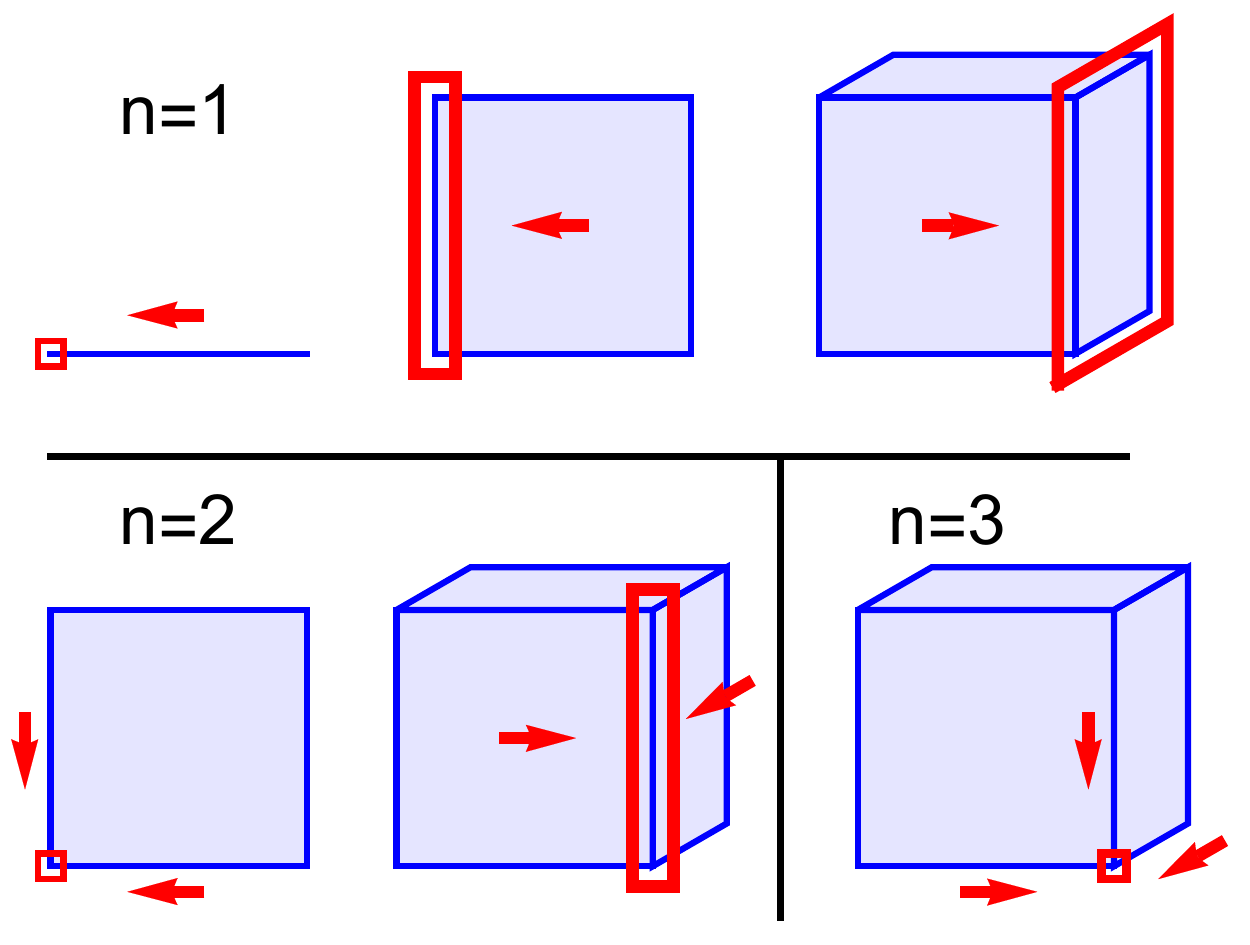}
\caption{Systems at the conventional ($\mdeg=1$ Maxwell critical point acquire zero modes (red) from missing bonds at open boundaries. Imbalances in these modes result from topological polarization (arrows) of the system's $\dim$-dimensional bulk. Systems with higher-order $\mdeg>1$ Maxwell rigidity have polarized surface elements, resulting in modes in corners and (in $\dim=3$) along 1D edges, defying conventional constraint counting.}
\label{fig:cartoon}
\end{figure}

\noindent For the case of $\mdeg=1$, this is simply Maxwell rigidity; more generally we refer to it as \emph{$\mdeg^{\textrm{th}}$-order Maxwell rigidity}. 
For $\mdeg = 1$, the constraint matrix resembles a non-Hermitian but square Hamiltonian, known to possess nontrivial topology~\cite{PhysRevB.84.205128,PhysRevA.87.012118,PhysRevLett.116.133903,PhysRevLett.118.040401,PhysRevB.95.174506,PhysRevLett.118.045701,PhysRevB.96.045437,PhysRevB.95.184306,xiong2018does,shen2018topological}, whereas for higher-order systems the constraint map connects spaces of modes and constraints that are different sizes.
For the present work we shall focus on the case $\dim=2,\mdeg=2$, in which zero modes are exponentially localized in corners. In 3D, one can have corner ($3^\textrm{rd}$-order) \emph{or} edge modes ($2^\textrm{nd}$-order rigidity), as indicated in Fig.~\ref{fig:cartoon}. Because of the duality between the rigidity and equilibrium matrices, it is also possible to have higher-order self stresses in \emph{under-constrained} systems, below the Maxwell point, satisfying $\mdeg = \nldof + 1 - \nlcon$. Bulk zero modes are always compatible with free boundary conditions; bulk self stresses are permitted by fixed boundaries.

\subsection{The higher-order topological invariant}

The topological paradigm is to relate the existence of boundary modes to bulk structure. We now describe how to relate the presence of the topological modes on our $(\dim-\mdeg)$-dimensional surface element to the topological degree of a map over the surrounding $(\dim-\mdeg+1)$-dimensional elements (e.g., a corner mode via the two adjoining edges). 

Let us describe a region $\reg$ of configuration space in which modes have amplitude $\uvec ^2 =1$ and $0 \le |z_i| \le 1, i = 1,2,\ldots n$ and are thus exponentially localized to the surface element in question. Consider then the nonlinear but continuous polynomial (or, in a more general gauge, Laurent polynomial) map from $\reg$ to the constraints $\evec = \rmat(\zvec)\uvec$.
 Given the criticality condition of Eq.~(\ref{eq:ourindex}), these spaces have the same dimension, and $\jac(\vvec)$, the Jacobian determinant of this map gives the signed ratio of their volume elements as a function of the vector coordinate $\vvec$.
As discussed in the Supplementary Material, the number of zeros in $\reg$ can then be determined by evaluating the number of times $\partial \reg$'s map to the normalized constraint space $\ehat = \evec/|\evec|$ covers the  real unit sphere $S^{2\nlcon-1}$:

\begin{align}
\label{eq:degree}
N_{\reg}=
\area_{2 \nlcon-1}^{-1}
 \int_{\partial \reg}d v_1 \ldots d v_{2 \nlcon-1} \jac(\vvec).
\end{align}

\noindent Here, $\area_j$ is the surface area of the $j$-dimensional hypersphere.  This topological map then relates the presence of zero modes in $\mdeg^\textrm{th}$ order critical systems to nonlinear maps from modes to constraints on the surrounding elements, such as the three edges adjoining the corner of a cube, generalizing the first-order topological invariant of Kane and Lubensky, a winding number that relates bulk structure to surface modes~\cite{kane2014topological}. Although the homotopy group of maps of spheres to themselves is integers ($\pi_n(S^n) = \mathbf{Z}$), our holomorphic maps always have nonnegative degree corresponding to the number of modes enclosed.

\subsection{Polarization of general surface elements}

We now discuss the mechanical polarization of structural elements. 
Consider a system with isolated zero modes, or zero modes which are isolated once transverse wavevectors are fixed (e.g., a 2D Maxwell structure has polarization which depends on transverse wavevector~\cite{rocklin2016mechanical}). One of its surface elements possesses a number of zero modes, which we refer to as its \emph{charge}, continuing terminology used in mechanics to describe not only topology~\cite{kane2014topological,paulose2015topological} but also purely geometrical singularities~\cite{moshe2018nonlinear}. Eq.~(\ref{eq:degree}) may be used to obtain this charge: for example, the charge on corners of a 2D second-order critical lattice are

\begin{align}
\label{eq:nll}
N_{\sigma_1,\sigma_2} = \int_{-1}^{1} d v_0\int_{-\pi}^{\pi} d v_1\,\int_{-\pi}^{\pi} d v_2\, \jac^{\sigma_1,\sigma_2}(\vvec)
\end{align}

\noindent where the arguments of the Jacobian describe real coordinates over values $z_i = r_i e^{i \theta_i}$ and, more generally, the mode shape $\uvec$. Since these are modes that bound the corner modes, they lie on the adjoining edges. To obtain in this manner the charge on, e.g., the upper left corner ($\sigma_1=-,\sigma_2 = +$), coordinates must be chosen to compactify the space and gauge must be chosen to maintain a holomorphic map as described in the Supplementary Material.

From  these charges we may define the polarizations of all elements of the structure. For example the second-order 2D structure has, in our language, polarizations of its bulk given by differences between charges of opposing edges and polarizations of its edges given by differences in corner charges. Generalizing this procedure, we may obtain higher-order polarizations, such as the quadrupole polarization of a 2D face of a 3D structure with third-order Maxwell rigidity.

This reveals nontrivial relationships between the charges and polarizations of various elements. The edge polarizations, along with the overall charge of the structure, suffice to determine the corner charges. However, the edge charges (or, alternately, the bulk dipole polarization) cannot determine the same without a topological quadrupole charge $q = \npp +\nmm - (\nmp + \npm)$ of the type described in higher-order electronic~\cite{benalcazar2017quantized,benalcazar2017electric,PhysRevLett.119.246402,PhysRevLett.119.246401,schindler2017higher} and phononic~\cite{serra2018observation}. Despite this intriguing connection, the present corner mode is clearly distinct from those quadrupole modes in that it occurs at zero frequency and doesn't require any mirror symmetries. Indeed, it applies even to systems with irregular boundaries.

While 0D corners necessarily have integer numbers of modes, 1D edges of 3D structures with second-order Maxwell rigidity may have fractional charge. Such a situation was observed in 2D~\cite{po2016phonon,rocklin2016mechanical} and 3D~\cite{baardink2017localizing} for Maxwell lattices. In our higher-order analog, we would expect the face between two fractionally-charged edges to host a sinusoidal zero-energy deformation.

\section{Model second-order system: the deformed checkerboard}
\label{sec:check}

\begin{figure}
\center
\includegraphics[scale=0.4]{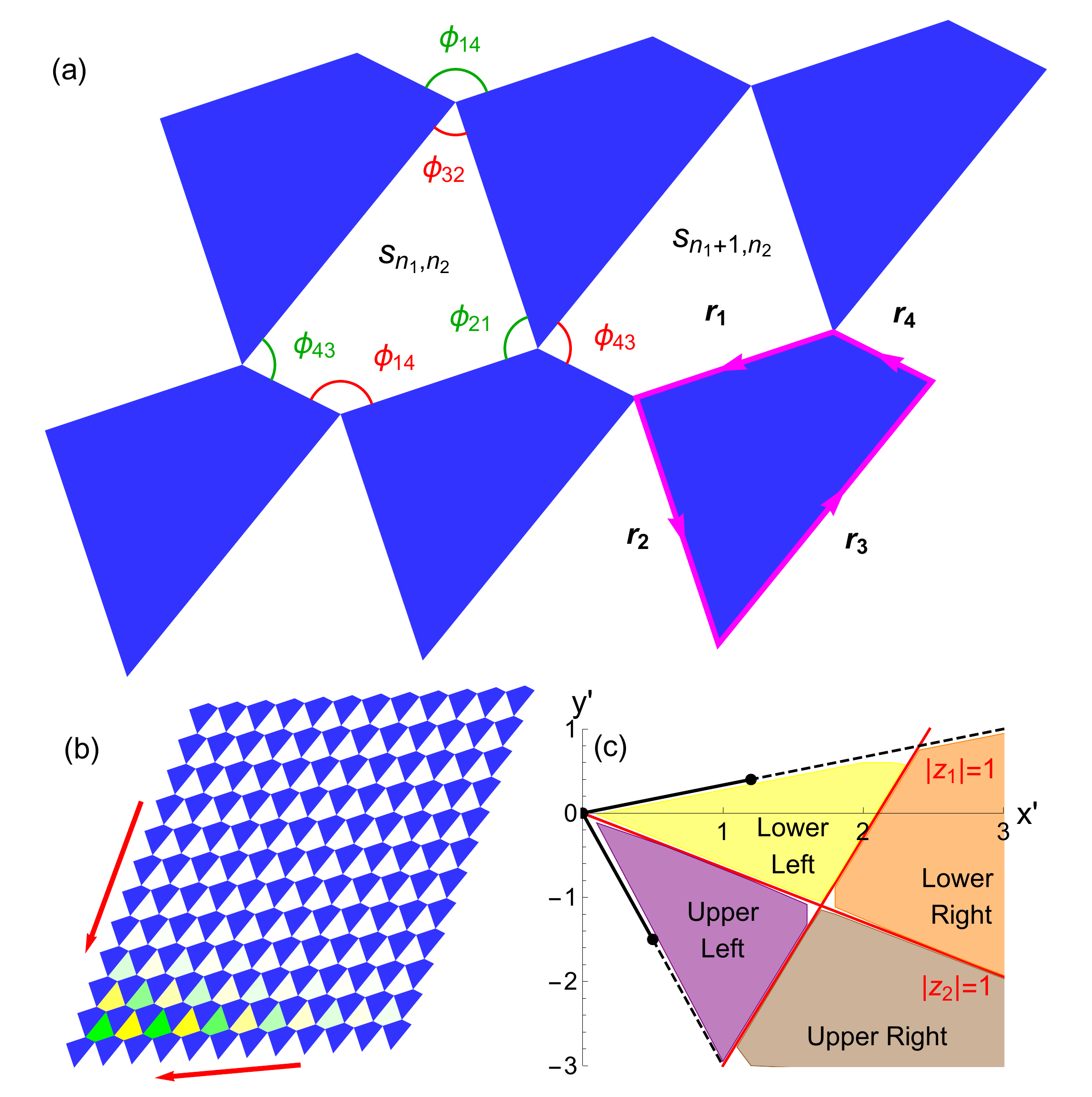}
\caption{(a) Solid pieces in the system are hinged and allowed to rotate relative to one another. A given void between pieces can shear, but this motion is coupled to that of four of its neighbors, leading to an overconstrained system. (b) The unique zero energy deformation is exponentially localized in one of four corners, with green and yellow shading indicating shearing in opposite directions. This topological mode lies between two topologically polarized edges (red arrows). (c) In a two-dimensional region of parameter space in which 
three of the piece's vertices are fixed and the fourth is placed at 
 $(x',y')$ the location of the corner mode is in agreement with the numerically obtained topological degree of the map (shading).
}
\label{fig:checkerboard}
\end{figure}

To model the general phenomenon of higher-order Maxwell rigidity, we consider the simplest case: second-order rigidity in 2D, which occurs in lattices with one additional constraint per cell beyond the Maxwell condition [Eq.~(\ref{eq:ourindex})]. Our chosen system is the deformed checkerboard, consisting of rigid quadrilaterals joined at free hinges as shown in Fig.~\ref{fig:checkerboard}, which can be thought of as the result of fusing two triangles together in the deformed kagome lattice~\cite{kane2014topological} or rigidifying an open quadrilateral in the deformed square lattice ~\cite{rocklin2016mechanical}. As shown in the Supplementary Material, any zero energy deformation of this system may be described by the scalar shearing of the voids between pieces of the form $s_{\ind_1,\ind_2}=s_0 z_1^{\ind_1}z_2^{\ind_2}$. Each void's shearing is coupled to that of its four neighbors by their shared vertices, resulting in the overconstrained constraint matrix

\begin{align}
\label{eq:rmatdef}
\rmat(z_1, z_2) &= \begin{pmatrix}
	b_1 + a_1 \, z_1 \\
	b_2 + a_2 \, z_2
         \end{pmatrix}.
\end{align}

\noindent As is now clear, the unique zero-energy deformation has $z_i = -b_i/a_i<0$, where the relative magnitudes of $a_i,b_i>0$ determine in which corner it is exponentially localized. This form reveals an important effect of symmetry: when $a_i=b_i$ and the constraint matrix is therefore invariant under under the reflection $\ind_i \rightarrow -\ind_i$ the mode lies on an edge rather than a corner. However, this symmetry is evident only in Eq.~(\ref{eq:rmatdef})---it corresponds to quadrilateral pieces that are not themselves symmetric but whose centers of mass lie on the lines connecting their opposing vertices. It is only when both such conditions are met, with parallelogram pieces, that the mode enters the bulk and extends to a nonlinear mechanism.

\subsection{Experimental Realization}

\begin{figure}
\center
\includegraphics[scale=0.38]{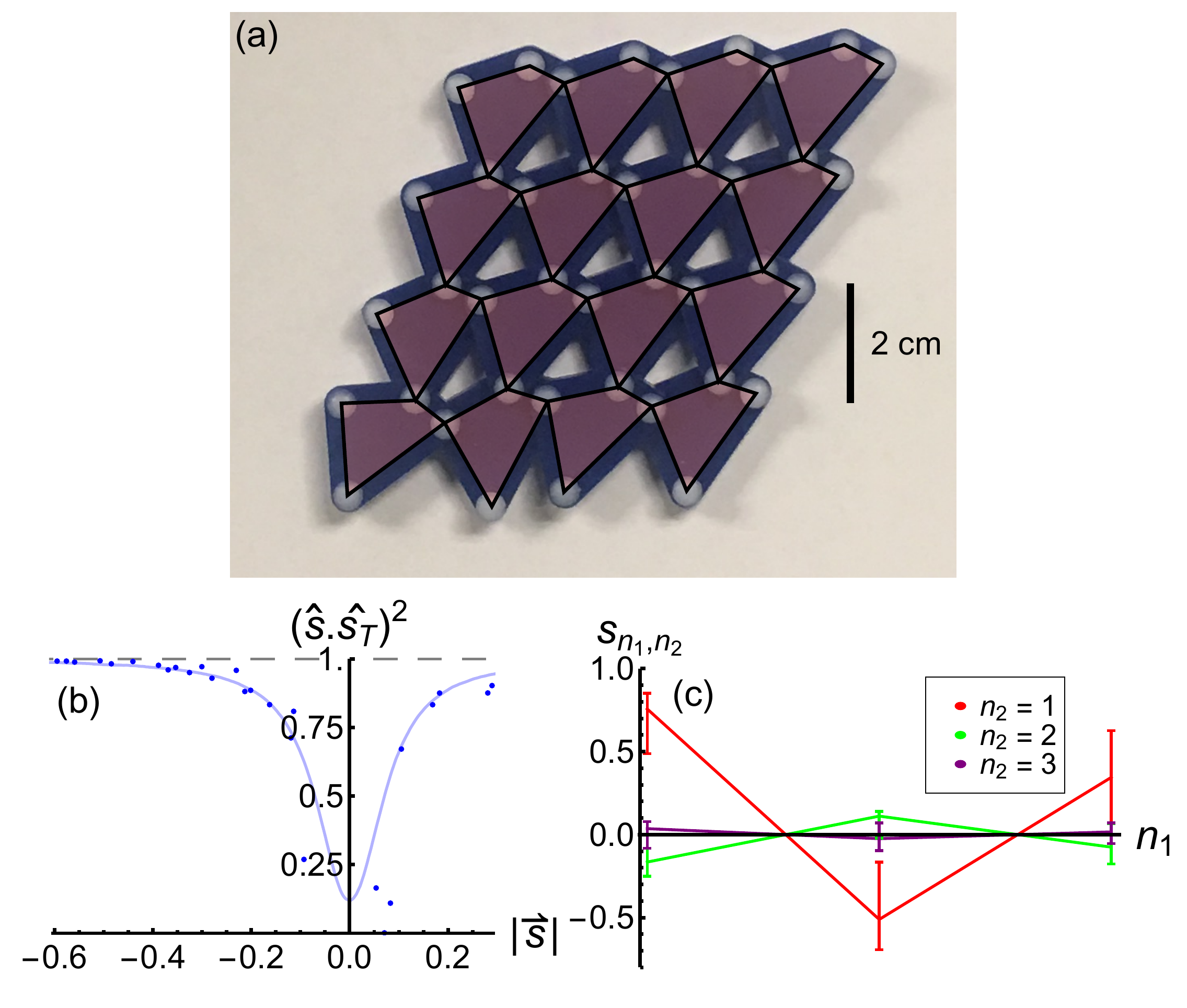}
\caption{(a) A hinged prototype of the checkerboard lattice. (b) Over a broad range of amplitudes, the topological component $\hat{s}_\textrm{T}$ is the dominant mode present. (c) Configurations in cells indexed by $(\ind_1,\ind_2)$, indicated by error bars, are in agreement with theory lines.}
\label{fig:experiment}
\end{figure}

We realize the topological metamaterial by using a high-precision programmable laser cutter (Trotec Speedy 300) to cut $\sim 1$cm pieces from 3.2mm thickness acrylic sheets. Nylon rivets are placed through snug holes in the pieces to join them at freely rotating hinges, with a $4 \times 4$ prototype then assembled, as shown in Fig.~\ref{fig:experiment}(a). The prototype is rigid throughout the bulk and most of the boundary, with a zero mode consisting of counter-rotating rigid pieces localized in the corner predicted by the constraints of Eq.~(\ref{eq:rmatdef}), as shown in Supplementary Video. 
This easily realized prototype permits the testing of the practical effects of friction, static disorder and geometrical nonlinearities on our idealized theory. These limitations prevent the system from acting as an exponential mechanical amplifier (as was recently treated at edges of disordered systems near the (first-order) Maxwell point~\cite{yan2017edge}) though some amplification in deformation is observed when the prototype is manipulated near the charged corner.

A digital camera is used to track the centers of the rivets as the system is deformed. Despite the linear nature of the theory, the compressibility of the rivets allows a $(-0.43,0.31)$ radians range (in the most-deformed void) of configurations in which static friction leaves the prototype stable without any external support. The vector of shears for the nine voids, $\vec{\sh}$, was tracked across this range, and compared to the predicted topological mode $\hat{\sh}_T$. The topological character, $\chart \equiv \left(\hat{\sh} \cdot \hat{\sh}_T\right)^2$ is plotted as a function of mode amplitude in Fig.~\ref{fig:experiment}(b), showing high overlap for all but the lowest-amplitude modes. The numerical prediction shown assumes a topological mode and independent normally-distributed errors throughout the system. As shown in Fig.~\ref{fig:experiment}(c), the topological mode is dominant everywhere save where its amplitude is lowest and static friction most relevant.

Thus, we have shown that even in real, easily realized systems possessing disorder, nonlinearities and friction the topological mode appears as predicted and accounts for a broad range of mechanical responses. 3D printing has already achieved 3D Maxwell systems~\cite{bilal2017intrinsically}, though a challenge remains in creating ``hinged'' pieces that rotate much more easily than they deform. Unlike 2D Maxwell lattices, which require careful control of the boundary because of the many deformation modes~\cite{rocklin2017transformable}, our system is mechanically stable because it has only a single, linear mechanism.

\section{Discussion}
\label{sec:discussion}

We have described a family of higher-order topological invariants that describes a class of periodic mechanical systems with zero-frequency boundary modes. These modes are protected not by symmetries but by a new index theorem relating the number of degrees of freedom and constraints to the dimensions of the bulk and boundary~[Eq.~(\ref{eq:ourindex})]. They are generated, protected, and placed on particular surface elements by the topological degree of the constraint map~[Eq.~(\ref{eq:degree})]. In particular, we have experimentally realized a two-dimensional structure with a single mode and demonstrated its mechanical response. The existence of the mode is protected by the index theorem, and its placement in a desired corner determined by a topological covering number, with further fine-tuning possible through geometric distortions. This allows for a material such as is realizable with 3D printing techniques~\cite{bilal2017intrinsically} with a unique programmed mode, topologically protected by its bulk structure.

This paradigm relies on the structure's periodicity and on particular boundary conditions, but is not limited to mechanical zero modes. In particular, their static counterparts, self stresses, have topological corner modes in under-constrained mechanical systems with fixed boundaries. Indeed, a Maxwell-Cremona dual~\cite{maxwell1864xlv,cremona1890two}, in which a mechanical network's vertices and faces exchange roles, exists for the deformed checkerboard with a topological corner self stress. In this way, both under- and over-constrained mechanical systems have topological modes. More generally, non-mechanical systems with varying numbers of constraints and degrees of freedom, such as spin systems~\cite{lawler2016supersymmetry} and electrical circuits~\cite{jia2013time,albert2015topological,mchugh2016topological} and others can have topological boundary modes protected by the index theorem and winding numbers.

Our systems lie at the intersection of two exciting areas of research. The first, Maxwell lattices with Kane Lubensky topological polarization, fall within our paradigm as systems with balanced numbers of degrees of freedom with boundary modes one dimension lower than their bulk. The original Maxwell index theorem [Eq.~(\ref{eq:index})] offers the advantage that the Maxwell modes extend nonlinearly, and exist despite disorder. In contrast, our more general modes are only linear (barring an additional symmetry, such as parallelogram tiles in the deformed checkerboard) and rely on perfect periodicity, though as our prototype demonstrates, it is still easy to realize the topological mode under realistic conditions. And such modes have the advantage that the rest of the structure is rigid, and that the modes are unique, rather than being part of a family that mix nonlinearly in ways that are difficult to control~\cite{rocklin2017transformable}.

The second area of research is into finite-frequency multipole modes~\cite{benalcazar2017quantized,benalcazar2017electric,PhysRevLett.119.246402,PhysRevLett.119.246401,schindler2017higher}, as exist at the corners of two-dimensional lattices. While these modes, including experimentally realized mechanical modes~\cite{serra2018observation} occur at finite frequency within band gaps and ours at zero frequency even in single-band systems, they seem to be related.
However, their origins seem fundamentally distinct: the finite-frequency systems have gapped edges owing to symmetry~\cite{benalcazar2017quantized,benalcazar2017electric,PhysRevLett.119.246402,PhysRevLett.119.246401,schindler2017higher}, while in the present study edges are gapped via an index theorem [Eq.~(\ref{eq:ourindex})] that counts their dimension. Indeed, acquiring (hidden) mirror symmetries actually closes the gap.

The topological connection between bulk and boundary, constraints and degrees of freedom, presents a number of immediate avenues for further study. Three-dimensional systems, which have already demonstrated unique features in Maxwell systems~\cite{baardink2017localizing}, should admit not only the 2D face modes of that study and the 0D corner modes corresponding to this one, but intermediate 1D edge (hinge) modes. Other systems, such as origami and kirigami~\cite{chen2016topological} have mixed dimension (a 2D sheet embedded in 3D space) or simply more intricate constraints~\cite{meeussen2016geared}. Finally, one may think of boundaries as a particular case of defects of given dimension, making contact with the extensive categorization of defects in topological insulators~\cite{teo2010topological}, admitting the same possibility of defect engineering observed in topological Maxwell lattices~\cite{paulose2015topological}.
Because of our map's nonlinearity, it may shed light on nonlinear excitations of polarized lattices~\cite{chen2014nonlinear}.

\emph{Acknowledgments--} The authors gratefully acknowledge helpful conversations with Bryan G. Chen, Michael Lawler, Tom Lubensky, Xiaoming Mao, Massimo Ruzzene, Christian Santangelo and Vincenzo Vitelli.

\bibliography{topomech}

\section*{Appendix 1: Constraint matrices}

\subsection{Constraints without periodicity}

Here we define a ``constraint matrix'', a linear map between a set of mode coordinates $\uvec$ and a constraint vector $\evec$. In a ball and spring system, these represent site displacements and bond extensions (positive or negative) respectively, but more generally modes in various systems can be described in terms of origami folding angles, potentials, currents, orientations of rigid bodies or rotors or the shearing motions used in the present work and described in detail in the following section. Regardless, the constraint map and an associated energy functional may be expressed as

\begin{align} 
\label{eq:cmat}
\evec &= \rmat \uvec, \\ \nonumber
\en &= \frac{1}{2} \evec^\textrm{T} \springs \evec \sim \frac{1}{2} \evec^2,
\end{align}

\noindent where zero-energy modes and modes in the nullspace of $\rmat$ coincide so long as $\springs$ is positive-definite. Since we are concerned purely with the zero-energy modes we can set $\springs$ to the identity matrix without effect. Then, from Eq.~(\ref{eq:cmat}b), it follows that the forces are

\begin{align}
\label{eq:qmat}
\fvec = - \frac{d \en}{d \uvec} = - \rmat^{\textrm{T}} \rmat \uvec = - \rmat^{\textrm{T}} \evec \equiv - \qmat \evec.
\end{align}

\noindent In a ball and spring system, these are forces on sites given by tensions in springs projected along the directions of springs. More generally, they are generalized forces resulting from the energy costs of violating the constraint equations. Continuing with the language of ball and spring systems, we introduce the equilibrium matrix, $\qmat \equiv \rmat^{\textrm{T}}$. Elements in its nullspace are referred to as self stresses, since the violations of the constraints generate stresses even in the absence of external force. It is the linear algebraic relationship between the constraint and equilibrium matrices that leads to the precise Maxwell index theorem given in the main text.

\subsection{Periodic systems with local constraints}

Consider, as in the main text, a periodically constrained system with crystal cells indexed by $\nvec = \left(\ind_1,\ind_2,\ldots \ind_d \right)$. The constraint equation then takes the form

\begin{align}
\label{eq:constraintperiodic}
\evec_{\nvec} = \sum_{\nvec'}\rmat_{\nvec - \nvec'} \uvec_{\nvec'},
\end{align}

\noindent where now $\uvec_{\nvec'}$ denotes the shape of the mode within cell $\nvec'$. This form of the constraint matrix, $\rmat_{\nvec,\nvec'} = \rmat_{\nvec - \nvec'}$ corresponds to systems invariant under the discrete translations $\nvec \rightarrow \nvec + \nvec^*$.
We then look for solutions of the form

\begin{align}
\label{eq:zperiodic}
\uvec_{\nvec} = \uvec \prod_{i} z_i^{n_i} \equiv \uvec \zvec^\nvec.
\end{align}

\noindent When $|z_i|=1$ these are simply Bloch wavefunctions, which span a finite system. However, in order to describe modes at the boundaries of large systems, it proves convenient to consider this broader family of modes. Indeed, when the logic of Bloch's theorem is repeated without any boundary conditions applied, modes of the above form are obtained.
This form similarly leads to constraints of the form $\evec_{\nvec} = \evec \zvec^\nvec$, and we thus convert Eq.~(\ref{eq:constraintperiodic}) to the form

\begin{align}
\label{eq:zconstraint}
\evec = \rmat(\zvec)\uvec.
\end{align}

\noindent Repeating this procedure for the equilibrium matrix, we obtain

\begin{align}
\label{eq:zeq}
\fvec &= \qmat(\zvec)\evec, \\ \nonumber
\qmat(\zvec) &\equiv \rmat^{\textrm{T}}(\zvec^{-1}).
\end{align}

For local interactions that are zero beyond some range smaller than the system size, the elements of $\rmat, \qmat$ are terminating Laurent polynomials in the several complex variables $z_i$. Because each bond is repeated periodically we have a gauge choice in whether to impose the constraint between, e.g., a particular cell and and its left neighbor or its right neighbor. We can use this constraint to convert the Laurent polynomial to a polynomial (i.e., multiply by powers of $z_i$ to remove poles without affecting zeros).

\subsection{Boundary conditions}

Solving $\rmat(\zvec) \uvec = 0$ ensures that the constraint equations are met in the bulk, but we are concerned particularly with modes with $|z_i| \ne 1$, which are exponentially localized to the boundary. We must then specify our boundary conditions. For zero modes, we choose free boundary conditions in which constraints that extend beyond the boundary are not present, ensuring that modes which satisfy the bulk constraints also satisfy the boundary conditions.

These boundaries need not be rectilinear. For example, a ``corner mode'' is present in systems with circular open boundaries. It is exponentially localized on a part of the boundary determined by $z_1,z_2$. However, we do not consider here interfaces between dissimilar regions.

For self-stresses, we choose instead fixed boundary conditions, so that degrees of freedom on the boundary are not in fact permitted to vary. This ensures that they do not move in response to unbalanced forces, again ensuring that a bulk mode satisfies the boundary conditions automatically.

\section{Constraint map}

\subsection{Configuration space}

In the previous section, we have treated the constraint matrix as a linear map between the vector $\uvec$ that describes the shape of a mode within a single cell and the vector $\evec$ of constraints. However, this matrix itself depends on $\zvec$, which describe how the mode varies between cells. Thus, we can also regard the \emph{constraint map} from the full space of modes to the constraint vector:

\begin{align}
\label{eq:conmap}
C:(\uvec,\zvec) \rightarrow \evec.
\end{align}

\noindent We concern ourselves with the zero modes, those that map to $\evec = 0$. A naive count would suggest that isolated zero modes should generically exist when the constraints are equal in number to the combined total number of $\uvec, \zvec$. However, this ignores the linearity of the results in $\uvec$, such that a nonzero scaling $\uvec \rightarrow \alpha \uvec$ never alters whether a mode satisfies the constraints. To reflect this, we restrict $\uvec$ to the complex projective space $C P^{\nldof-1}$, while leaving $\zvec,\evec$ to lie in the full spaces of $\ind,\nlcon$ respective complex variables.

\subsection{Counting argument}

Each constraint can be thought of as a hyperplane two dimensions smaller passing through the full mode space $(\uvec, \zvec)$. Zero modes lie in the intersections of these several hyperplanes. Some constraints can be either redundant (two constraints specifying that an angle assume the same value) or incompatible (two constraints requiring that a single angle assume two distinct values). Excepting these non-generic cases, we expect the dimension of the space of zero modes to be reduced by two for each additional complex constraint. 
This leads then to the counting argument presented in the main text [Eq.~(\ref{eq:ourindex})] for $\mdeg$, the number of dimensions lower the topologically charged surface elements are than the system itself
(e.g., $\mdeg = 2$ for modes localized to 0D corners in 2D structures):

\begin{align}
\mdeg = \nlcon + 1 - \nldof.
\end{align}

\noindent The additional term $1$ comes because linearity in $\uvec$ reduces the dimension of the space of zero modes as discussed above. Note that the $\zvec$ contribute only $\ind \le \dim$ complex dimensions because we choose to hold fixed the remaining $\zvec$.

\subsection{Choice of gauge}

The periodicity of the bulk of a system ensures that when a constraint is satisfied in one cell it is satisfied in all cells. This gives a choice of which constraint we wish to satisfy. For example, in a 1D system with a constraint connecting neighboring cells, one could choose either to satisfy the constraint as it exists between the origin cell and its right neighbor or between the origin cell and its left neighbor, or even the constraint as it exists between degrees of freedom, e.g., seven and eight cells to the right of the origin. All choices are equally valid, but not equally useful. We select for our choice of gauge the \emph{minimal holomorphic} gauge.

Consider a constraint that involves degrees of freedom in some cells indexed by $\{\nvec_i\}$. We can impose the gauge shift $\nvec_i \rightarrow \nvec_i+\nvec_g$. We choose the unique vectorial cell index $\nvec_g$ such that each element of each cell index is non-negative. In this way, our constraint map will include only nonnegative powers of $\zvec$, making it holomorphic in  $(\uvec,\zvec)$, a property that we exploit later in arriving at our application of a topological degree theorem. This gauge removes all poles (in which a constraint diverges) while avoiding introducing spurious zeros (in which the constraint map is satisfied without satisfying the physical constraints).

\subsection{Topological degree theorem}

Let us define the notion of the \emph{degree} of a map, a concept well known in mathematical topology. In particular, following Flanders Chp. 6.2~\cite{flanders1963differential}, the degree of a map $f$ from a closed, oriented $(n-1)$-dimensional manifold $\partial \mathbf{M}$ to the surface of the unit hypersphere $\mathbf{S}^{n-1}$ is the (signed) integer number of times that hypersphere is covered:

\begin{align}
\label{eq:degdef}
\textrm{deg} f = \frac{1}{A_{n-1}} \int_\mathbf{\partial M} f^* \sigma',
\end{align}

\noindent where $A_{n-1}$ is the surface area of $\mathbf{S}^{n-1}$ and $\sigma'$ is on $\partial \mathbf{M}$. This degree is invariant under deformations of the manifold that do not cross zeros of $f$. This is homotopy invariance, and indeed complete homotopy invariance in which two maps from the unit sphere to itself are homotopic to one another if and only if they have the same degree.

Proceeding from this deformability in the manner of Gauss's law, the surface $\partial \mathbf{M}$ may be decomposed into infinitesimal surfaces surrounding each isolated zero in the region $\mathbf{M}$ bounded by $\partial \mathbf{M}$. Hence, the degree of the map is equal to the sum of the degrees of each zero. A nonzero degree then indicates the presence of at least one zero (corresponding to a physical mode) in the region. Limiting this is that since these zeros can have positive or negative degrees themselves, a zero degree does not rule out the possibility of having multiple zeros (some with positive and some with negative degrees) in the region of interest. Hence, for real-valued mappings topological degree as measured through the above integral serves as a powerful but incomplete tool for identifying topological modes.

However, as discussed above, our class of systems can be described (in the correct gauge) via complex, holomorphic functions that introduce additional structure much as analyticity permits residue theory for single-variable complex functions. D'Angelo, Chp. 
2~\cite{d1993several} considers a holomorphic map from the boundary of a space of $n$ complex coordinates to the equal-dimension unit sphere, $\mathbf{S}^{2 n - 1}$ and likewise concludes that the degree is the sum of the degrees of over isolated zeros within the region. The holomorphic case differs, however, in that now these degrees are always positive (the Jacobian of the map is real and positive). Hence, for our holomorphic case the degree of the map identifies the number of zeros in the region of interest, weighted by their positive integer degrees (or ``indices''), a type of multiplicity.

In the following subsection, we will show that the Jacobian of our particular system, which involves not only a complex space but a complex projective space, is similarly positive. Taking that for now as a given, we obtain the crucial result of the main text:

\begin{align}
\label{eq:degreesupp}
N_{\reg}=
\area_{2 \nlcon-1}^{-1}
 \int_{\partial \reg}d v_1 \ldots d v_{2 \nlcon-1} \jac(\vvec).
\end{align}

That is, for some region of interest $\reg$ in the space of linear modes $(\uvec,\zvec)$, the number of zero modes present in the region (counting multiplicity) is equal to the number of times the constraint map takes the boundary $\partial \reg$ to the unit hypersphere. In particular, the number of modes present in a corner of a 2D structure (e.g., $|z_1|<1$ and $|z_2|<1$) is equal to the number of times the constraint map takes the modes on the adjoining edges ($|z_1|<1$ and $|z_2|=1$ or $|z_2|<1$ and $|z_1|=1$) to the unit sphere.

In order for the result to hold, $\reg$ must be compact. Thus, regions with, e.g., $|z_1|>1$, are not covered. However, any given corner of the system may be considered separately by relabeling the index so that, in this case, as $n_1$ increases one moves from right to left rather than left to right. Thus, the total charge on an edge may be obtained by summing over different charges, obtained in different coordinate systems and gauges. Alternately, if it's known that all zeros occur in some compact region, then the region $\reg$ may be inflated and the limiting value of the degree is the number of zeros in the non-compact region.

\subsection{Complex projective space and positive degree}

As discussed above, the vector $\uvec$ describing the shape of a mode is within a cell is treated as being in complex projective space. That is because our constraint map is linear in $\uvec$ (but not $\zvec$), so that $\uvec \rightarrow \lambda \uvec$ does not change whether a mode is a zero mode. However, the above result that all zeros have positive degree relies on the positivity of the Jacobian of the constraint map. This condition holds for holomorphic maps between complex spaces, but it is not obvious that it extends to maps from complex projective spaces. Here we present a simple argument demonstrating just that.

Consider a related constraint map, 

\begin{align}
\label{eq:conmaptwo}
C:(\uvec',\zvec) \rightarrow (\evec,e').
\end{align}

\noindent Here, $\uvec' \in \mathcal{C}^{\nldof}$ is simply a complex vector (except that we exclude $\uvec'=0$). The additional constraint $e' \equiv \uvec'\cdot \lc - 1=0$ is then necessary to prevent $\uvec \rightarrow \lambda \uvec$ (which would violate it for $\lambda \ne 1$). We could simply stop here, and use this holomorphic map between two topological spaces equivalent to $\mathbf{S}^{2 \nlcon + 1}$ to determine the degree. Instead, we reduce this to the complex projective space that proves most convenient, at least for the present application. Consider a set of coordinates for $\uvec'$ $\{\pvec,\uvec'\cdot \lc\}$, where the $\phi_j$ are $2 (\nldof-1)$ real coordinates describing the position of $\uvec$ in complex projective space. The Jacobian of the full map then becomes

\begin{align}
\label{eq:fulljac}
\begin{pmatrix}
    \frac{\partial \evec}{\partial \zvec}       & \frac{\partial \evec}{\partial \pvec}  &
		\frac{\partial \evec}{\partial \uvec' \cdot \lc } \\
        \frac{\partial e'}{\partial \zvec}       & \frac{\partial e'}{\partial \pvec}  &
		\frac{\partial e'}{\partial \uvec' \cdot \lc }
\end{pmatrix}.
\end{align}

\noindent Recognizing that the bottom row is simply $(0,\ldots 0, \mathcal{I})$, this Jacobian has the same determinant as that of 
$(\frac{\partial \evec}{\partial \zvec}, \frac{\partial \evec}{\partial \pvec})$, the original constraint map. Hence, since the large Jacobian has a positive real determinant the smaller one does as well. The one remaining concern is that when $\uvec' \cdot \lc = 0$ the coordinates $\pvec$ are ill-defined. However, the general topological result permits deformation of the manifold into small neighborhoods of the isolated zeros, and so to show that the Jacobian is positive there it only remains to choose some $\lc$ such that this condition is not met in those neighborhoods.

\section{The Deformed Checkerboard Lattice}

\subsection{Constraint geometry}

As described in the main text, our model system consists of corner-sharing quadrilaterals that are rigid but allowed to rotate freely against one another, as shown in Fig.~\ref{fig:checkerboard}(a). Our counting argument indicates that such a system may have some number of linear zero modes beyond global rigid-body motions (two translations and a rotation). We wish then to choose a set of coordinates that captures any potential zero mode and a set of constraints on them that enforce the physical requirement that the pieces themselves not deform. Note in particular that because we are not concerned with distinguishing between finite-energy configurations that violate the constraints, our choice of coordinates need not capture configurations that obviously violate the constraints, such as those which alter the distance between two portions of a particular piece. To illustrate the fact that the topological result does not depend on a particular choice of coordinates, we describe three such choices, selecting the one which renders further calculations most straightforward.

The first such method is to employ a conventional compatibility matrix. Since each vertex of the quadrilateral piece is shared between two pieces, two vertices are needed per unit cell, and are each permitted to move in two dimensions. To render the quadrilateral piece rigid, five spring constraints must be used, e.g., along the four edges and across one of the diagonals. Thus, by this count the system has one more constraint than degree of freedom per cell. This leads to a compatibility matrix with five rows and four columns, which unnecessarily complicates both the calculations and the analytical theory.

\begin{figure}
\center
\includegraphics[scale=0.38]{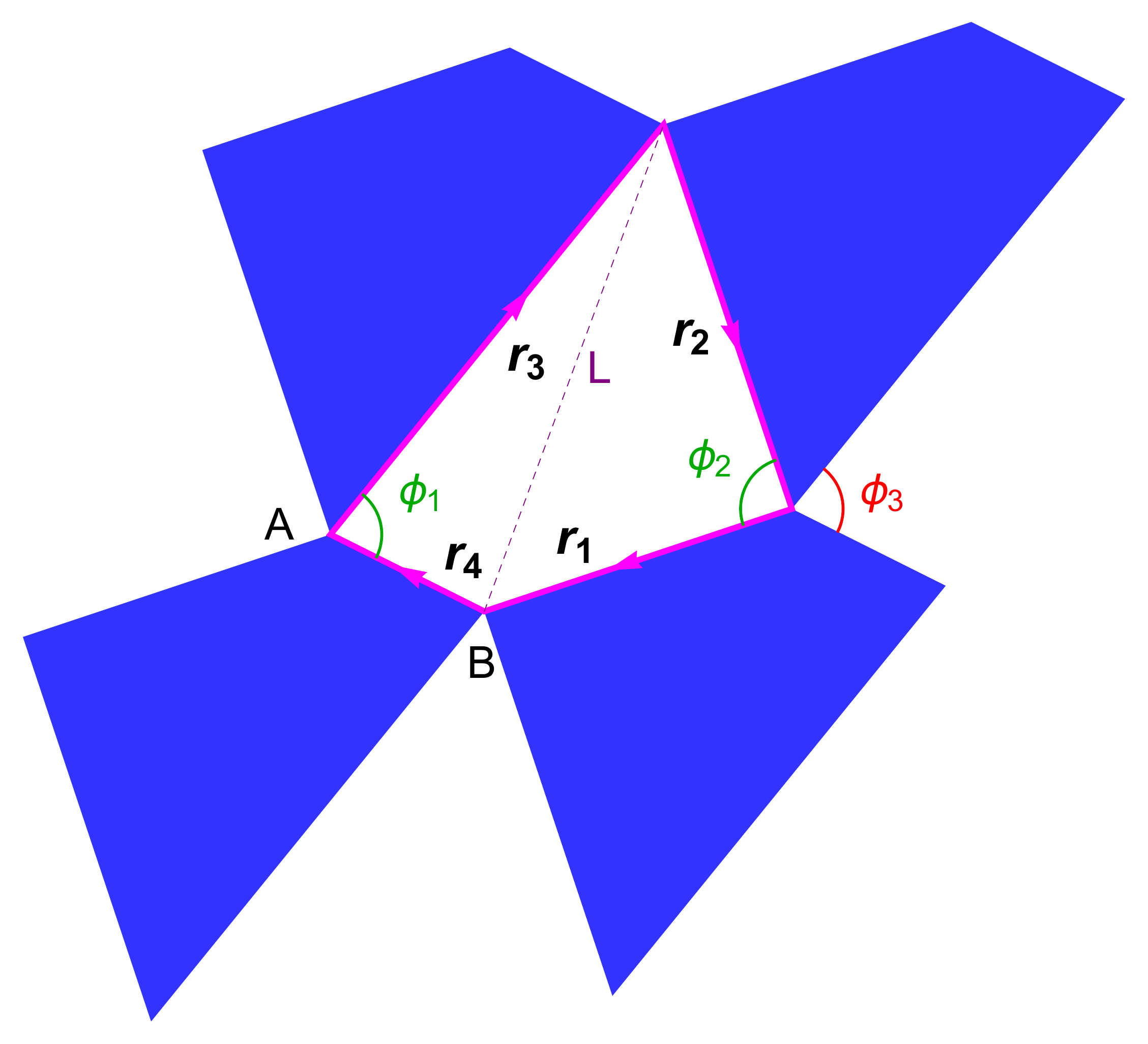}
\caption{Representation of a $2 \times 2$ checkerboard pieces. The vertices $A, B$ along the 5 connecting edges form the unit cell in the sites and bonds characterization. The angles $\phi_1, \phi_2, \phi_3$ represent the shearing motions across each void.}
\label{fig:checkerrep}
\end{figure}

A second choice is to take advantage of the fact that we know any valid zero mode will permit only translations and rotations of the pieces. It thus follows that if the vectors along each of the four edges of a \emph{void} between four quadrilateral pieces sum to zero then the configuration corresponds to a valid zero mode. In fact, we can ignore translations altogether, relying upon the fact that the vertices of two adjoining pieces will remain in contact for a valid zero mode. Then, we can parametrize a system by the vectors $\ra, \rb, \rc, \rd$ along the edges of a piece (and summing to zero) and a potential zero mode by the angles of rotation $\rotm_{\ind_1, \ind_2}$ of the pieces. Our constraint then becomes (see Fig.~\ref{fig:checkerrep}):

\begin{align}
\label{eq:rotcon}
\rd \theta_{n_1, n_2} + \rc \theta_{n_1, n_2 + 1} +
\rb \theta_{n_1 + 1, n_2 + 1} + \ra \theta_{n_1 + 1, n_2} = 0.
\end{align}

\noindent Assuming the z-periodic form $\rotm{\ind_1, \ind_2} = \rotm_0 \, z_1^{\ind_1} \, z_2^{\ind_2}$, this requires

\begin{align}
\label{eq:rotconz}
\rd  + z_1 \ra +  &z_2 \rc + z_1 z_2 \rb  = 0.
\end{align}

By this formulation, we have reduced the shape of the mode within the cell to a single complex number $\rotm_0$ and, in fact, because our constraint is linear not even this matters. This simplifies a problem that initially had a four-dimensional space of modes and five constraints to one with two constraints and only a trivial mode shape.
The only coordinates which matter are $z_1,z_2$, which describe the spatial variation of the mode between cells. While this dramatically reduces the dimensionality of the problem, Eq.~(\ref{eq:rotconz}) couples the two lattice directions in a way that, we shall see, is not essential.

Our third and preferred way of enforcing the constraints is to note not only that a mode is valid when the voids between cells close but to note that since they consist of quadrilaterals, they have only a single floppy shearing motion. Characterizing the strength of the shearing in each void suffices to entirely determine the configuration of the pieces, up to overall translations and rotations. Consider, as shown in Fig.~\ref{fig:checkerrep}, the dependence of the length $L$ of the void diagonal on two of its interior angles:
\begin{align}
\label{eq:length1}
L^2 = r_3^2 + r_4^2 - 2 r_3 r_4 \cos (\ta), \nonumber \\
L^2 = r_1^2 + r_2^2 - 2 r_1 r_2 \cos (\tb).
\end{align}

\noindent From this, and noting that $\partial \tc/\partial \tb = -1$, we have

\begin{align}
\frac{\partial \tc}{\partial \ta} &= - \left ( \frac{\partial L^2}{\partial \ta} \right ) \,  \left( \frac{\partial L^2}{\partial \tb} \right)^{-1} \\ \nonumber
&= - \frac{r_3 r_4}{r_1 r_2} \frac{\sin \ta}{\sin \tb}.
\end{align}

\noindent Treating the two-dimensional vectors as being embedded in three dimensions, we can write:
\begin{align}
\label{eq:partial}
\frac{\partial \tc}{\partial \ta} =-  \frac{|\rc \times \rd|}{|\ra \times \rb|}.
\end{align}

\noindent
Assuming as before that we have some mode $s$ that shears these voids with the extent $s_{n_1, n_2} = s_0 \, z_1^{n_1} \, z_2^{n_2}$, we recognize that $\tc = \ta z_1$, with similar behavior in the second lattice direction. Linearizing the constraints, we have then

\begin{align}
\label{eq:shearcon}
|\rc \times \rd| + |\ra \times \rb|z_1 = 0, \nonumber \\
|\ra \times \rd| + |\rc \times \rb|z_2 = 0.
\end{align}

\noindent Now, we immediately obtain that the zero mode satisfying the constraints  of a system parametrized by $\{\mathbf{r}_i\}$ appears at

\begin{align}
\label{eq:shearz}
z_1=-\frac{|\rc \times \rd|}{|\ra \times \rb|}, \nonumber \\
z_2=-\frac{|\ra \times \rd|} {|\rc \times \rb|}.
\end{align}

These results reveal an idiosyncratic symmetry of our system. Note that, for example, $|\ra \times \rb|$ is the area of the triangular portion of the quadrilateral portion to the left of the diagonal. The zero modes thus follow the mass: if a quadrilateral has most of its area to the left of its vertical diagonal and below its horizontal diagonal then its zero mode lies in the lower left-hand corner, etc. The system becomes symmetric under $\ind_1 \rightarrow -\ind_1$ only when the piece's center of mass lies on the vertical diagonal. In this case the system's sole floppy zero mode lies on the lower edge. Thus, the condition that generates symmetric constraint equations is not a conventional spatial symmetry of the pieces themselves.

\subsection{Numerical calculation of topological degree}

The constraints of Eq.~(\ref{eq:shearcon}) are of the form of the constraint map of Eq.~(\ref{eq:rmatdef}):

\begin{align}
\label{eq:dccon}
\fvec = (f_1, f_2) &= (b_1 + a_1 z_1, b_2 + a_2 z_2).
\end{align}

\noindent Because we have only a single complex degree of freedom per cell, $s$, which we can set to 1 because of linearity, our space of modes is simply determined by $(z_1,z_2)$. The region corresponding to modes exponentially localized to the lower left-hand corner and its boundary are, respectively,

\begin{align}
\label{eq:llregion}
\reg_{\textrm{LL}} &= \{|z_1| \leq 1, |z_2| \leq 1 \} \\\nonumber
\partial \reg_{\textrm{LL}} &= \{|z_1| \leq 1, |z_2| = 1 \} \cup \{|z_1| = 1, |z_2| \leq 1 \} 
\end{align}

\begin{figure}
\center
\includegraphics[width=.45\textwidth]{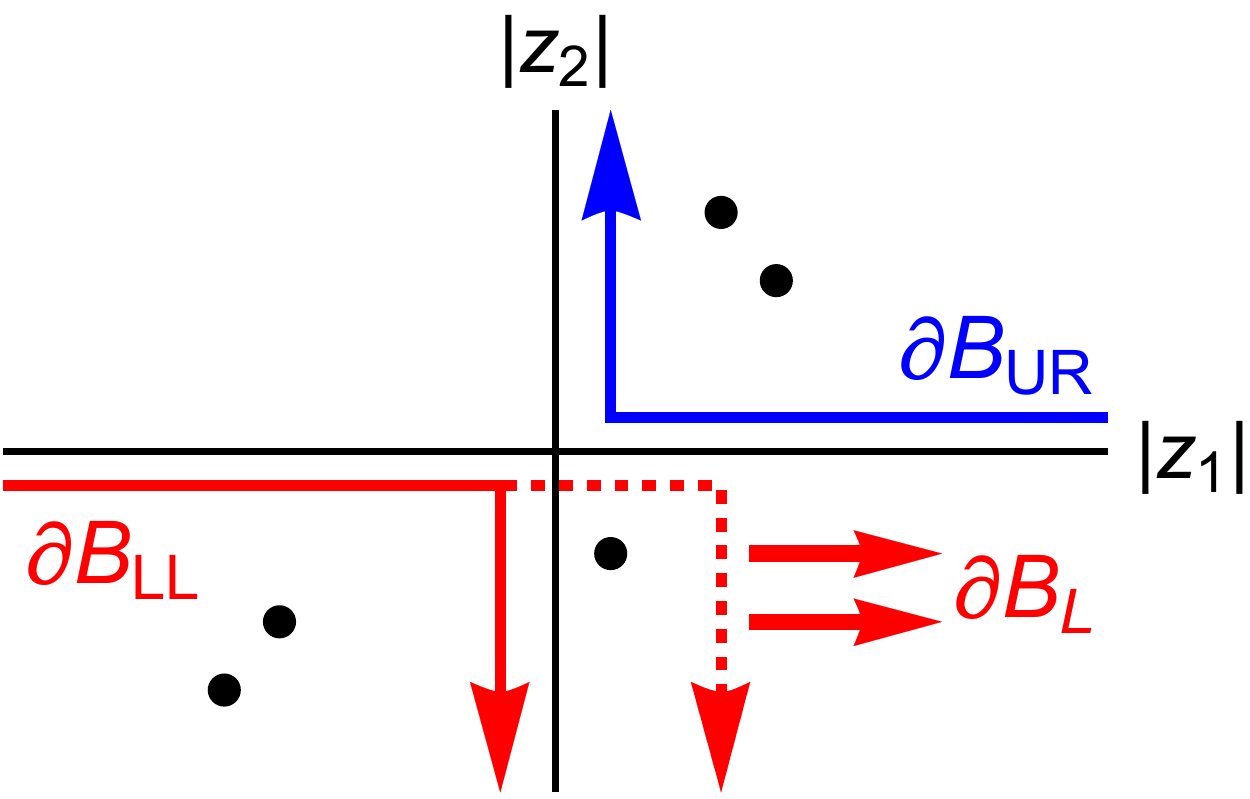}
\caption{Zero modes of the deformed checkerboard lattice are shown as points at given mode coordinates $|z_1|,|z_2|$, with the phases of $z_i$ not shown. Modes exponentially localized in the lower-left corner of the system appear in the lower-left corner of the plot, with the boundary of this region marked by a red arrow. This region may be extended to the right (dashed arrow) to capture the number of zero modes (charge) in total between the two corners, with care taken with this non-compact region. Similarly, to obtain the number of zeros in the upper-right corner, an integral is taken over a surface in a different gauge, as indicated by the blue arrow.}
\label{fig:intreg}
\end{figure}

\noindent This region is shown as a red arrow in Fig.~\ref{fig:intreg}. In order to perform the integral over it, we need to parametrize our three-dimensional surface $\partial \reg$, which we do as:

\begin{align}
\label{eq:integralparm}
(z_1,z_2)=
\begin{cases} 
\left((1+v_0) e^{i v_1},e^{i v_2}\right)
&\mbox{if } v_0 < 0 \\ 
\left(e^{i v_1},(1-v_0) e^{i v_2}\right)
&\mbox{if } v_0 \ge 0 \end{cases} \\
-1 \le v_0 \le 1, \nonumber \\
-\pi \le v_1 \le \pi, \nonumber \\
-\pi \le v_2 \le \pi, \nonumber
\end{align}

\nonumber Although our underlying space is complex, we wish similarly to express our constraints in terms of real numbers, such that $\fvec \equiv (\textrm{Re} f_1,\textrm{Im} f_1,\textrm{Re} f_2,\textrm{Im} f_2)$ and $\hat{f} \equiv \fvec / |\fvec|$. In determining the volume element of the Jacobian map from $\partial \reg$ to the three-sphere we encounter the minor issue that $\fvec$ lies in a three-dimensional space embedded in a four-dimensional one. We can obtain the volume in three-dimensional space most easily by including a fourth vector that is orthonormal to all $\partial_{v_i} \hatf$. $\hatf$ itself serves this purpose, leading to a Jacobian volume density described in terms of column vectors as

\begin{align}
\label{eq:jaccol}
J(\vvec) = \det \left[ \hatf, \partial_{v_0} \hatf,\partial_{v_1} \hatf,\partial_{v_2} \hatf\right].
\end{align}

\noindent Hence, the degree of the map, which gives the number of zeros in $\reg$, is for a given checkerboard lattice

\begin{align}
\label{eq:intex}
N_\reg = \area_{3}^{-1}
 \int_{-1}^{1} d v_0\int_{-\pi}^{\pi} d v_1\,\int_{-\pi}^{\pi} d v_2\,  d v_0 d v_1 d v_2 \, J(\vvec).
\end{align}

\noindent This result depends, via the constraint map, on the particular vectors $\mathbf{r}_i$ along the edges of the checkerboard piece. As seen in Fig.~\ref{fig:checkerboard}(c), the numerical calculation of the topological degree agrees with the direct calculation for the lower-left corner. However, to obtain the charges on the remaining corners care must be taken with the indexing and choice of gauge.

\subsection{Charges on additional corners}

Although they seem as readily addressed as the lower-left corner, modes localized to the lower right corner  have $|z_1|>1$, meaning that in our original coordinate system $\reg_{\textrm{LR}}$ is not compact, invalidating our topological theorems. This is easily resolved by choosing a coordinate system such that the index $n_1$ assumes its lowest value on the right edge and counts up as one moves leftward. This creates a coordinate system in which $z_1 \rightarrow 1/z_1$ and $\reg_{\textrm{LR}}$ becomes compact.

However, this takes the constraint $b_1 + a_1 z_1 = 0$ to $b_1 + a_1/z_1 = 0$. In order to apply the theorem, we must recover a minimally holomorphic gauge by scaling the constraint by $z_1$. We then recover the old result with $a_1 \leftrightarrow b_1$. Thus, by relabeling our system and making the correct choice of gauge, we can obtain the number of zero modes in each corner. In this language, the most readily obtained charge, that of the lower left hand corner, is called $N_{--}$, with the other corners labeled $N_{\pm\pm}$.

In contrast, if we attempt to find the total number of zero modes on the lower edge, a region we can label $\reg_{\textrm{L}}$, we find that there is no gauge in which this region is compact. We can obtain the total charge on such regions by summing over the corners involved, or by extending the region $\reg_{\textrm{LL}}$ to $\reg_{\textrm{L}}$ as shown in Fig.~\ref{fig:intreg}. This method obtains the number of zeros on the lower edge as the limit over compact regions, permissible when all zeros are known to lie in a compact region, as is the case in generic physical systems of the sort considered here.

\end{document}